\def\aj{AJ}%
\def\araa{ARA\&A}%
\def\apj{ApJ}%
\def\apjl{ApJ}%
\def\apjs{ApJS}%
\def\aap{A\&A}%
\def\aaps{A\&AS}%
\def\mnras{MNRAS}%
\def\pasp{PASP}%
\newcommand{\kms}{\mbox{km$\,$s$^{-1}$}}

\newcommand{\msun}{\mbox{${\rm M}_{\odot}$}}

\newcommand{\dr}{\mbox{${\rm d}$}}

\newcommand{\tcr}{\mbox{$t_{\rm cr}$}}

\newcommand{\dndt}{\mbox{$\dr N/\dr t$}}
\newcommand{\rsmc}{\mbox{$R_{\rm SMC}$}}

\newcommand{\spaze}{\hspace{0.19cm}}

\documentclass[useAMS,usenatbib]{mn2e}

\usepackage{amssymb,latexsym,graphicx,natbib,times}

\title[Stellar structure in the SMC]
  {Evolution\spaze of\spaze stellar\spaze structure\spaze in\spaze the\spaze Small\spaze Magellanic\spaze Cloud}

\author[M. Gieles et al.]
  {M.~Gieles,$^{1}${\bf\footnote{}} N.~Bastian$^{2,3}$ and B.~Ercolano$^{2,4}$\\
  $^1$ European Southern Observatory, Casilla 19001, Santiago 19, Chile \\
$^2$ Institute of Astronomy, University of Cambridge, Madingley Road, Cambridge,
 CB3 0HA, UK \\
    $^3$ Department of Physics and Astronomy, University College London, Gower Street, London, WC1CE 6BT, UK\\
      $^4$ Harvard-Smithsonian Center for Astrophysics, 60 Garden Street, Cambridge, MA 02138, USA
}
\date{Accepted 2008 September 12. Received 2008 September 11; in original form 2008 May 29}

\pagerange{\pageref{firstpage}--\pageref{lastpage}} \pubyear{2008}

\def\LaTeX{L\kern-.36em\raise.3ex\hbox{a}\kern-.15em
    T\kern-.1667em\lower.7ex\hbox{E}\kern-.125emX}

\begin{document}         

\maketitle

   \begin{abstract} 
   The projected distribution of stars in the Small Magellanic Cloud (SMC) from the Magellanic Clouds Photometric Survey is analysed.
  Stars of different ages are selected via criteria based on $V$ magnitude and $V-I$ colour, and the degree of  `grouping' as a function of age is studied.  We quantify the degree of structure using the two-point correlation function and a method based on the Minimum Spanning Tree and  find that the overall structure of the SMC is evolving from a high degree of sub-structure at young ages ($\sim$10\,Myr) to a smooth radial density profile. This transition is gradual and at $\sim$75\,Myr the distribution is statistically indistinguishable from the background SMC distribution. This time-scale corresponds to approximately the dynamical crossing time of stars in the SMC. The spatial positions of the star clusters in the SMC show a similar evolution of spatial distribution with age. Our analysis suggests that stars form with a high degree of (fractal) sub-structure, probably imprinted by the turbulent nature of the gas from which they form, which is erased by random motions in the galactic potential on a time-scale of a galactic crossing time.
\end{abstract}
\begin{keywords}
galaxies: star clusters -- galaxies: Magellanic Cloud -- galaxies: structure 
\end{keywords}

\section{Introduction}
\label{sec:intro}
The\blfootnote{{\bf$^\star$}E-mail: mgieles@eso.org} majority of stars form in clustered environments (e.g. \citealt*{2000prpl.conf..151C} and \citealt{2003ARA&A..41...57L}, henceforth LL03), but only a small fraction ($\lesssim10$ per cent) of stars ends up in bound star clusters which survive the embedded phase.  It was noted much earlier that almost all stars in the Galactic disc form in (unbound) OB associations  (e.g. \citealt{1957PASP...69...59R, 1979ApJS...41..513M}) and that residual gas expulsion is the most probable explanation (e.g. \citealt*{1980ApJ...235..986H, 1984ApJ...285..141L}).
LL03 introduced the term `infant mortality' for the rapid dissolution of embedded clusters at young (few Myrs) ages.  

More recently, the infant mortality scenario has also been used to explain the strong drop in the age distribution, \dndt, around $\sim$10-20\,Myr of clusters in the Antennae galaxies and M51 (\citealt*{2005ApJ...631L.133F} and  \citealt{2005A&A...431..905B}, respectively).  \citet{2007ApJ...658L..87P} introduced a new approach for studying infant mortality, by comparing the spatial distribution of stars of different stellar types in the nearby spiral galaxy NGC~1313. They found that O-type stars are more strongly grouped than B-type stars, which is what is expected if all stars form in clusters and the majority disperses on time-scales comparable to the life-time of O-stars (few Myrs).

The term infant mortality is now used in widely varying contexts. First, the time-scales involved range from $~3\,$Myr (LL03) up to a Gyr \citep{2005ApJ...631L.133F,2006ApJ...650L.111C}. Second, the term `star cluster' is used for irregular, dynamically unmixed, groups of embedded stars with total masses of $10-100\,\msun$ (e.g.~LL03), for compact  (few pc) and massive (few times $10^5-10^6\,\msun$) clusters in interacting galaxies (e.g.~\citealt{1996AJ....112.1839S}) and for  large scale ($\gtrsim100\,$pc) stellar associations \citep{2007ApJ...658L..87P}.

The Small Magellanic Cloud (SMC) cluster system has been heavily under debate recently. \citet{2005AJ....129.2701R} presented a new catalogue of SMC star clusters and showed the existence of a declining \dndt. This was  interpreted as infant mortality working on a 3 Gyr time-scale by \citet{2006ApJ...650L.111C}.  However, \citet*{2007ApJ...668..268G} showed that the decline  is in fact due to detection incompleteness at higher ages,  later confirmed by an independent analysis \citep{2008MNRAS.383.1000D}.

This study aims at tying the above issues together to give a more coherent picture of the evolution of structures in the SMC. 
Inspired by the work of \citet{2007ApJ...658L..87P}, the catalogue of five million stars detected in the Magellanic Clouds Photometric Survey (MCPS) \citep{2002AJ....123..855Z} is used to select stars of different ages. In section~\ref{sec:data} the data and the selection criteria of different regions are described.
In section~\ref{sec:clustering} two techniques are employed in order to search for stellar groups and quantify their evolution with age.
The implications and conclusion are presented in  section~\ref{sec:conclusions}. 

\section{Data and selection criteria}
\label{sec:data}

%

\begin{figure}
	\begin{center}
  \includegraphics[width=8cm]{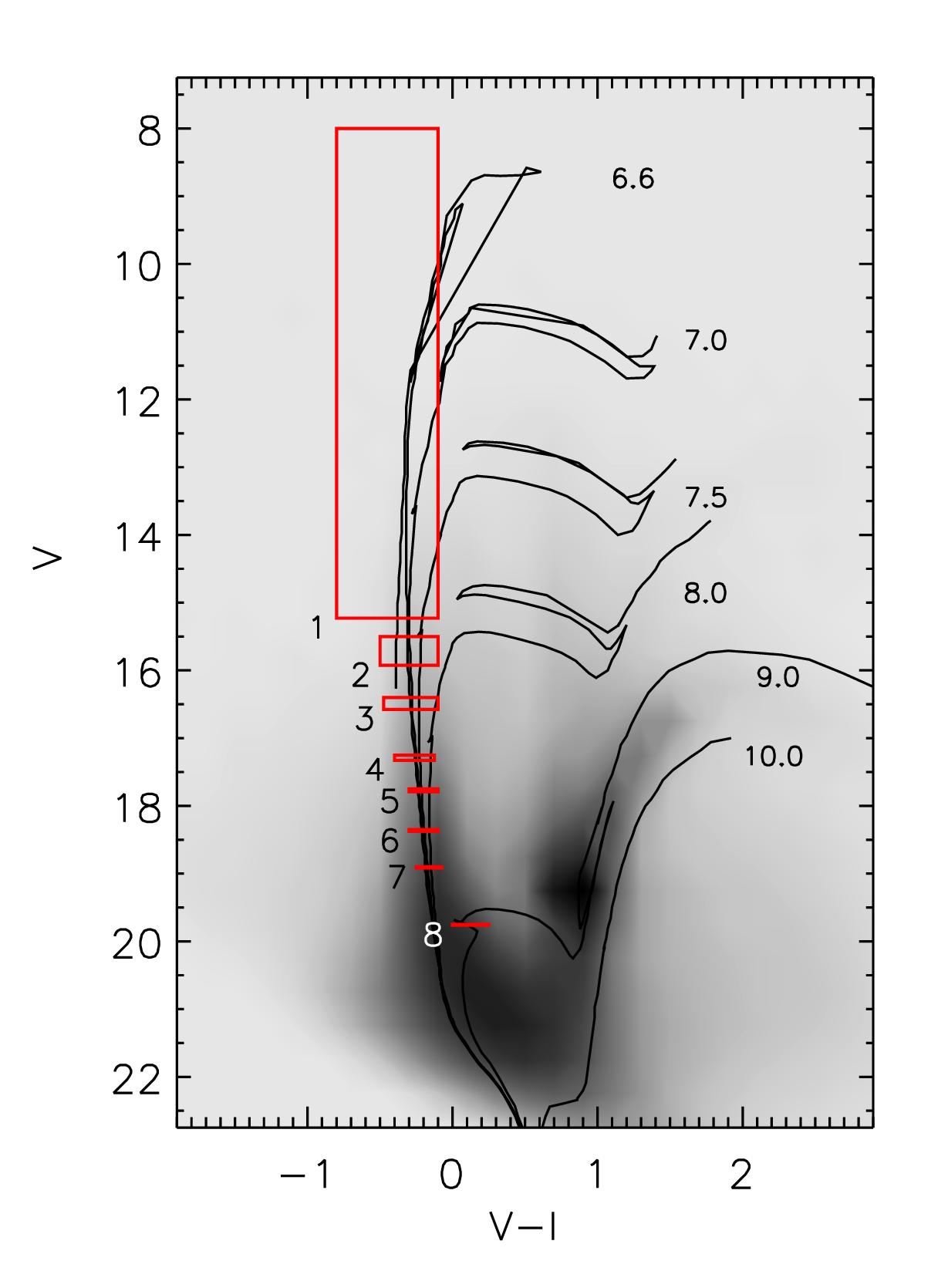}
  \end{center}
  \vspace{-0.05cm}
       \caption{Colour-magnitude diagram for stars from the catalogue of \citet{2002AJ....123..855Z}. 
       The shading represents the number of stars present (shown as the square-root to increase the contrast), with dark representing a higher number of stars.  The selected boxes are indicated.  The numbers on the right-hand side of the panel refer to the base-ten logarithm of the age of the isochrones shown.  
 }
    \label{fig:cmd}
\end{figure}

The MCPS covers an area of roughly $4.^\circ5\times4^\circ$ of the Small Magellanic Cloud (SMC). We use the Johnson $V$ and the Gunn $I$ photometric results  of roughly five million stars to construct the colour magnitude diagram (CMD) of $V$ vs. $V-I$. The photometry was corrected for a Galactic foreground extinction of $A_V=0.12\,$ mag \citep*{1998ApJ...500..525S}. The resulting CMD is shown in Fig.~\ref{fig:cmd}, where the greyscale represents the number of stars with that colour and magnitude (dark representing higher number).
Overplotted are  evolutionary isochrones of the Padova models for $Z=0.004$ \citep{1994A&AS..106..275B,1996A&AS..117..113G,2000A&AS..141..371G} which were converted to the $UBVRIJHK$ photometric system by \citet{2002A&A...391..195G}. We adopted a distance modulus of 18.87 mag ($\sim$60\,kpc) \citep*{2003MNRAS.339..157H}.

Based on the CMD we select seven non-overlapping boxes around the main sequence and one box around the  turn-off of a one Gyr population, each containing 1500 stars, in a circle with a diameter of 2.7 degrees around the mean RA and DEC of the stars in the catalogue.  This avoids the outer regions of the galaxy which have a higher fraction of contaminating sources.   We use equal numbers of stars to avoid possible number dependancies in the analysis (section~\ref{sec:clustering}). The box sizes are determined as follows: once criteria for the $V-I$ colour and peak $V$ magnitude are chosen, the stars satisfying these constraints are ordered in $V$ brightness and the brightest 1500 are selected. The magnitude of the faintest star included in the final 1500 stars sample defines the lower boundary of the box.  The boxes around the main sequence always sample stars of multiple ages, with the age spread getting larger for boxes lower in the main sequence. This means that even though the mean age increases going down in the main sequence, we always have a contribution of the youngest population.
Box~8 samples  only turn-off stars with ages of roughly a Gyr.

\begin{figure*}
\begin{center}
  \includegraphics[width=17.cm]{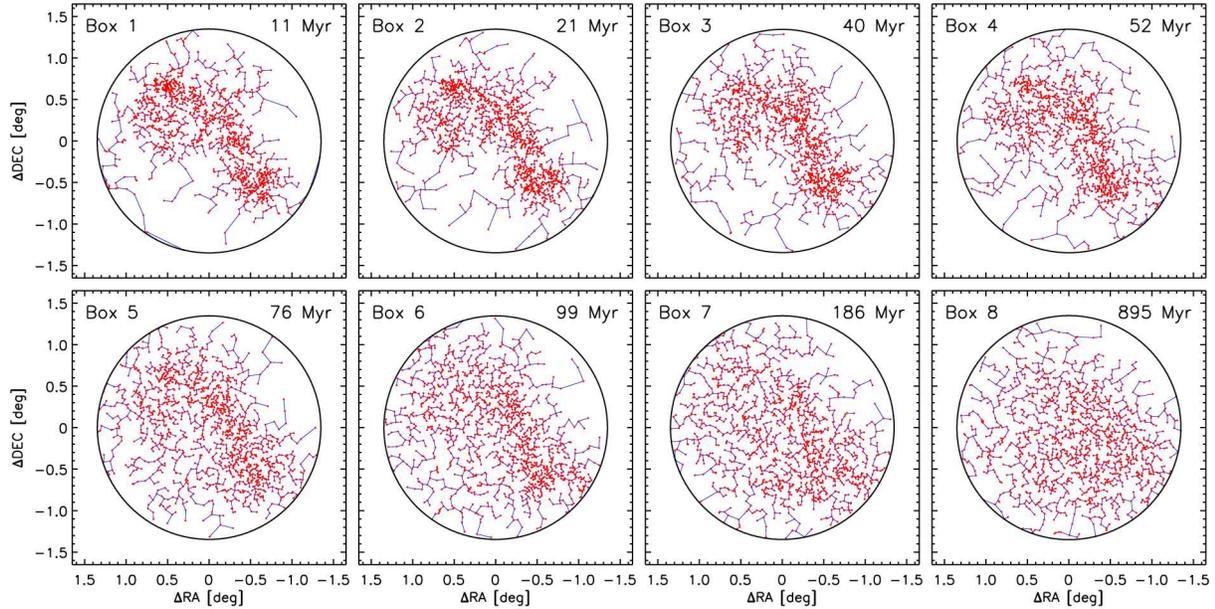}
\end{center}

       \caption{In each panel we show the RA and DEC (relative to the mean) of the 1500 stars in each age box as dots.  The mean age of the stars in the boxes (section~\ref{sec:data}) is indicated in each panel. The Minimum Spanning Trees  (MSTs, see section~\ref{ssec:mst}), of each sample is over-plotted with lines. }

    \label{fig:mst}
\end{figure*}
\begin{figure}
   \includegraphics[width=8.cm]{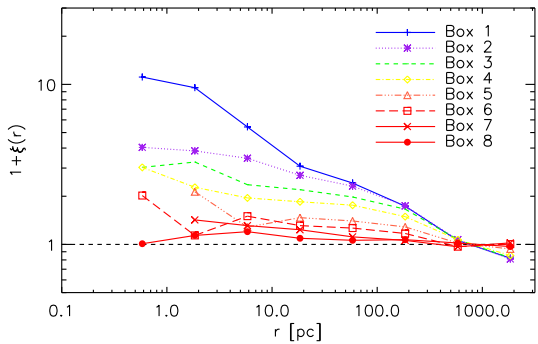}
       \caption{The two-point correlation function (TPCF) for stars in the eight age boxes.}

    \label{fig:tpcf}
\end{figure}
\begin{figure}
\begin{center}
   \includegraphics[width=8.cm]{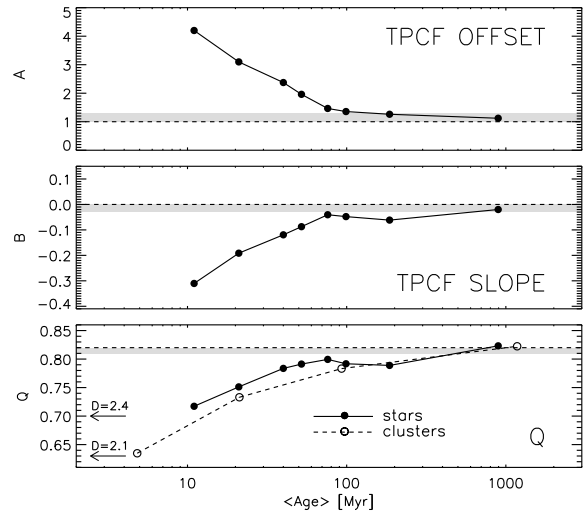}
\end{center}
       \caption{Evolution of the three structure indicators discussed in section~\ref{sec:clustering} as a function of mean age of the eight boxes.  In the top and middle panels the results of the power-law fits (equation~\ref{eq:tpcf})  to the TPCFs are shown. In the bottom panel the results for $Q$ of the clusters(stars) are shown as open(filled) circles. The  dashed lines indicate the value corresponding to the reference distribution. The grey areas denote the values where the difference between the value for the youngest age box and the reference value is less than 10 per cent.  }  
   \label{fig:panel}
\end{figure}

To quantify the contribution of the different ages to the boxes, we perform the following test: we assume a constant rate of star formation over the last Gyr (older ages do not contribute to the selected boxes on the main sequence, see Fig.~\ref{fig:cmd}) and assign masses according to a Salpeter mass function. 
We assign colours and magnitudes to each star, based on their age and mass as derived from the Padova models and the adopted distance modulus.  For each age we find the number of stars with colour and magnitude satisfying  our criteria, allowing us to derive the mean age of stars in each box.  See \citet{2007MNRAS.379.1302B,bastianetal08} for details.

We do not take into account internal extinction variations in the SMC. \citet{2004AJ....127.1531H} find a trend of younger stellar population ($\lesssim10\,$Myr) being more extincted than older stellar populations ($\gtrsim1\,$Gyr). Especially for box~1 this  could result in target stars being outside the box due to extinction,  but we do not expect many sources to be shifted into our selection boxes. For the most extincted field in the sample of \citet{2004AJ....127.1531H}, the $A_V$ distribution of the young population peaks around $0.5\,$mag, corresponding to $E(V-I)\simeq0.2$ mag.  For most of their fields, however,  the extinction for the young population is much lower.

In Table~\ref{tab:boxes} we summarise the selection criteria of the eight age boxes and the resulting  minimum, maximum and  mean ages of the stars in each box.   In figure~\ref{fig:mst} we show with points the distribution of the 1500 stars in each age box. The lines show the Minimum Spanning Trees (MSTs) derived from the distribution of the stars  (section~\ref{ssec:mst}).

In section~\ref{sec:clustering} we will   use the two-point correlation function and the statistical $Q$-parameter  \citep{2004MNRAS.348..589C} that can be derived from the MST, in order to search for structure evolution in the SMC.   We compare the distribution of stars (and star clusters) of different ages to a distribution without sub-structure, which we will refer to as the {\it reference distribution}.  This distribution has a smooth power-law density profile, $\rho(r)$,  with $r$ the distance to the centre of the SMC and $\rho(r)\propto r^{-0.75}$. Such a  profile, projected in 2D, is a reasonable approximation of the observed distribution of stars with $V<20$.

\section{Structure evolution in the SMC}
\label{sec:clustering}

\subsection{The two-point correlation function}
\label{ssec:tpcf}

A commonly used and relatively straightforward technique to measure sub-structure is the two-point correlation function (TPCF), which has been used extensively as a statistical measure of large-scale cosmological structure (e.g.~\citealt{1980lssu.book.....P}).  
We define the TPCF, $\xi(r)$, as

\begin{equation}
1+\xi(r)=\sum_{i=1}^N n_i(r){\mbox{\LARGE{/}}}\sum_{i=1}^N n_i^{\rm ref}(r),
\end{equation}
where $n_i$ is the number of stars found at a distance $r$ from star $i$, and $n_i^{\rm ref}$ is the same, but for  the references distribution. For the sum over the reference distribution we take the average of 100 realisations. This is similar to the procedure implemented by  \citet{1993AJ....105.1927G} who studied the spatial distribution of young stellar objects in the Taurus-Auriga molecular cloud.

\begin{table*}
\begin{center}
\caption{Overview of the selection criteria of  the eight boxes shown in Fig.~\ref{fig:cmd}. }
\label{tab:boxes}                                              
\begin{tabular}{lrrrrrrr}
\hline
   &\multicolumn{2}{c}{$V-I$}   \hspace{-0.35cm}   &  \multicolumn{2}{c}{$V$} &\multicolumn{3}{c}{Age [Myr]}\\
Box\hspace{-0.3cm} & Minimum & Maximum  & Minimum &Maximum &Minimum&Maximum& Mean \\ \hline
 1  &$ -0.800$  &$ -0.100$  &$ 8.00$  &$15.23$  &$4$  &$25$  &$11$ \\
 2  &$ -0.500$  &$ -0.100$  &$15.50$  &$15.93$  &$4$  &$40$  &$21$ \\
 3  &$ -0.475$  &$ -0.100$  &$16.40$  &$16.58$  &$4$  &$79$  &$40$ \\
 4  &$ -0.400$  &$ -0.125$  &$17.25$  &$17.33$  &$4$  &$126$  &$52$ \\
 5  &$ -0.300$  &$ -0.100$  &$17.75$  &$17.79$  &$4$  &$200$  &$76$ \\
 6  &$ -0.300$  &$ -0.100$  &$18.35$  &$18.38$  &$4$  &$200$  &$99$ \\
 7  &$ -0.250$  &$ -0.075$  &$18.90$  &$18.92$  &$4$  &$316$  &$186$ \\
 8  &$  0.000$  &$  0.250$  &$19.75$  &$19.76$  &$794$  &$1000$  &$895$ \\\hline
\end{tabular}
 \end{center}
\end{table*}

The results for  all boxes are shown in Fig.~\ref{fig:tpcf}.  The slope gets shallower with increasing age (box number) and approaches that of the reference distribution, that is, $1+\xi(r)=1$.   We fit power-law functions of the form 

\begin{equation}
1+\xi(r)=A\,\left(\frac{r}{10\,{\rm pc}}\right)^B
\label{eq:tpcf}
\end{equation}
 and follow 
 the evolution of the parameters $A$ and $B$ from equation~(\ref{eq:tpcf}) with age for the eight age boxes. This is shown in the top and middle panels of Fig.~\ref{fig:panel}.
The values evolve very quickly towards the reference values ($A=1$ and $B=0$).  For the stars in box~5 (with a mean age of $\sim$75\,Myr) the difference between the initial value and the reference value has already reduced by 90 pre cent for both $A$ and $B$. This simple analysis indicates that sub-structure in the SMC is erased in $\sim$75\,Myr.


\subsection{The Minimum Spanning Tree and the $Q$-parameter}
\label{ssec:mst}


In order to quantitatively study the distribution of groups of sources objective algorithms must be developed.  Additionally, large datasets, such as the one used in the present work, require high levels of automation.  One such tool which has been developed and successfully employed in the study of young star-forming regions is the Minimum Spanning Tree (MST) algorithm \citep{1991A&A...244...69B}.  All points, that is, the positions of the sources, in the distribution are connected to their nearest neighbour, in order to form a single `tree', such that the total length of all the connecting segments (edges) is minimised and no closed loops are formed.
This is a  mathematically unique way of describing a dataset.  In Fig.~\ref{fig:mst} we show the MSTs for the stars in the eight age boxes.  

The MST can be used to study the distribution of stars and quantify the amount of sub-structure, as was done by e.g.~\citet{2004MNRAS.348..589C} and \citet{2006A&A...449..151S} who applied the MST method to study structure in  (embedded) Galactic clusters.

The $Q$-parameter, which was introduced by \citet{2004MNRAS.348..589C}, is a measure of the amount of sub-structure within a spatial distribution. In short, the $Q$-parameter is defined as $Q\equiv\bar{m}/\bar{s}$, with $\bar{m}$ the normalised mean edge length of the MST and $\bar{s}$ the normalised mean correlation length.  The normalisations are applied to make the result independent of the size and number of sources in the distribution.

\citet{2004MNRAS.348..589C} showed that $Q$ is a unique way of distinguishing between smooth radial density gradients and (fractal) sub-clustering. 
For a random 3D distributions of stars,  projected in 2D,  the value of $Q$ is 0.79. For distributions with more sub-structure $Q<0.79$, and for smooth density profiles which are centrally concentrated, $Q>0.79$ (see their Fig.~5).  
We estimate $Q$ for the reference distribution through interpolation of the values given in Table~1 in \citet{2004MNRAS.348..589C} and find $Q_{\rm ref}\simeq0.82$.

We determined $Q$ for the stellar distribution in the eight age boxes and find an increase of $Q$ with age (see filled circles in the bottom panel of Fig.~\ref{fig:panel}). We also determine $Q$ for the star clusters in four age bins containing equal number of clusters. The cluster positions and ages are taken from the catalogue of \citet{2003AJ....126.1836H}. The result is shown as open circles. The $Q$ results for the stars and the clusters agree very well in the age range where they overlap. The ages of the clusters are known with higher accuracy than the ages of the stars, for which we have stars with a large age spread contributing to boxes~1 to 7. Therefore, the agreement between the $Q$ values shows that the ages assigned by our selection method (section~\ref{sec:data}) are likely to be representative.  The offset to slightly higher $Q$-values for the stars  is probably due to the fact that the stars in each box, contrary to the clusters, have always stars of the youngest age contributing. It indicates that the distribution of the stars older than the mean age affect the value of $Q$ more than the distribution of the younger stars. This is confirmed by simulations of stellar distributions  \citep{bastianetal08}. 

The arrows in the bottom panel of Fig.~\ref{fig:panel}  indicate the results for a (projected)  fractal model with dimension 2.1 and 2.4 \citep{2004MNRAS.348..589C}.
 We find that the cluster and stellar distributions in the young age boxes have a lot of sub-structure, with box~1 showing the same $Q$ value as a fractal with dimension $\sim$2.4.  This probably reflects  the structure of the turbulent gas from which they are formed \citep{2001AJ....121.1507E}.   For increasing age box, $Q$ increases rapidly, levelling off  around box~5 ($\sim$75\,Myr), where the value is  within 10-20 pre cent that of the reference distribution.
This is in agreement with the time-scale indicated by the TPCF. It suggests that sub-structure originating from the star formation
process is erased within $\sim$75\,Myr.

\subsection{Sensitivity of the methods}
Both the TPCF and the $Q$-parameter indicate that for stars in boxes 5 and higher the distribution is essentially indistinguishable  from the reference distribution. From a visual inspection of Fig.~\ref{fig:mst} we see that the distribution of stars is still evolving going from box~5 to box~8, while according to the TPCF and the $Q$-parameter these distributions are statistically comparable (Fig.~\ref{fig:panel}).
This could indicate that both our methods are insensitive to structure that is present only on a low signal-to-noise level. However, we note that the main body of the SMC also affects our results. In box~5 we can clearly recognise the shape of the SMC, while this is less evident in box~8. This structure, which is of galactic scale, will increase both the value of $Q$ and the slope of the TPCF. So, if the average distribution of stars is somewhat more centrally concentrated than our reference distribution, this could compensate the effect small structures have, such that the values are similar to those of the reference distribution. We emphasise that the absolute values of both methods are sensitive to edge effects and to some degree suffer from degeneracy when the distribution is a mixture of a pure fractal and a centrally concentrated distribution. This leads us to conclude that the variations on the 10-20 per cent level should not be over-interpreted. However, the general evolution of, and the agreement between, both the TPCF structure indicators and the $Q$-parameter is a strong indication that stellar structure is erased on a short   ($<100\,$Myr) time-scale in the SMC.

\section{Conclusions }
\label{sec:conclusions}
We have studied the evolution of sub-structure in the distributions of stars of different ages in the Small Magellanic Cloud (SMC). Using two different methods, namely the two-point correlation function (TPCF), and the statistical $Q$-parameter, we find that the stellar distribution evolves to the   reference distribution, that is, a smooth power-law density profile,  on a time-scale of $\sim$75\,Myr.

Our results qualitatively agree with what  \citet{2007ApJ...658L..87P} found for stars in NGC~1313. However, we have also quantitatively determined the time-scale in which `stellar groupings' disappear ($\sim$75\,Myr). In the scenario of \citet{2007ApJ...658L..87P} all stars form in clusters/groups, which dissolve due to gas expulsion. The super-virial velocities of stars after gas removal are of the order of a few $\kms$. Combined with the radius of the SMC galaxy ($\rsmc\simeq2\,$kpc), the time it would take for stars that travel from a dissolving cluster to cover one galactic radius is $\sim$1\,Gyr,  which is much longer than the 75\,Myr we find. 
\citet{2008MNRAS.386..826E} measured a velocity dispersion, $\sigma$, of $30\,\kms$ for (massive) field stars in the SMC, which combined with the radius of the SMC results in a dynamical crossing time, $\tcr=\rsmc/\sigma\simeq75\,$Myr. This is the same as the time-scale for structure dispersal that we found in our study, making random motions in the galactic potential a plausible explanation for the erasure of structure.

 \citet{2008A&A...482..165G} showed that only a small fraction ($\sim$3 per cent) of stars that are formed in the SMC end up in  clusters that survive past the embedded phase, i.e. those that can be identified as genuine star clusters in optical studies. Whether this is due to a 97 per cent infant mortality rate of embedded clusters or due to a large fraction of stars forming in the field, can not be judged from the current data.  We do know that already at very young ages ($\sim$10\,Myr) the majority of stars are not in dense clusters and these `field stars'  still show the  fractal imprints of the natal gas, which is being erased dynamically  by random motions in the galaxy potential. 

In an accompanying paper \citep{bastianetal08} we further develop and explain the techniques and conclusions reported here, and compare these results to those found for the Large Magellanic Cloud.

\section*{Acknowledgement}
We thank an anonymous referee for constructive comments that helped to improve this paper.

\bibliographystyle{mn2e}

\end{document}